\documentclass[10pt,twocolumn]{article}

\usepackage[margin=0.75in]{geometry}
\usepackage{amsmath,amssymb}
\usepackage{graphicx}
\usepackage{booktabs}
\usepackage{siunitx}
\usepackage{subcaption}
\PassOptionsToPackage{hyphens}{url} % allow long bibliography URLs to break at hyphens
\usepackage[colorlinks=true,linkcolor=black,citecolor=blue!65!black,urlcolor=blue!65!black]{hyperref}
\usepackage{cite}
\usepackage[T1]{fontenc}
\usepackage{circuitikz}
\usepackage{tikz}
\usetikzlibrary{arrows.meta,positioning,shapes.geometric}
\usepackage{balance} % balance columns on last page

\graphicspath{{figures/}}
\title{\textbf{Physics-to-Circuit Analysis of GaN RF Integrated Circuits\\
versus GaAs and Silicon}}

\author{
\normalsize Rajesh Vedala (Microsoft, Surface Team)\\[1pt]
\normalsize Palak Kapoor, Harkirat Kaur (NIT Warangal, Department of Physics)\\[3pt]
\small\textit{This work was carried out during August 2020; manuscript
updated and drafted in February 2026.}
}
\date{}

\begin{document}
\maketitle

\begin{abstract}
\noindent The migration of radio-frequency (RF) integrated-circuit
platforms from silicon to GaAs and now to gallium nitride is derived
here from first principles. The hexagonal non-centrosymmetric GaN
lattice admits a macroscopic polarization; elasticity and the
piezoelectric tensor fix the bound sheet charge at an AlGaN/GaN
interface, and Poisson's equation with triangular-well quantisation
yields a degenerate quasi-two-dimensional channel of
$\sim\!10^{13}\,\si{cm^{-2}}$ with no doping. Energy--momentum
conservation for pair creation and phonon-limited energy relaxation set
the breakdown field (\SI{3.3}{MV/cm}) and saturation velocity
($2.5\times10^{7}\,\si{cm/s}$), which combine into geometry-free
limits $V_\mathrm{br}f_T=E_c v_\mathrm{sat}/2\pi$ and
$R_\mathrm{on}^\mathrm{sp}=4V_\mathrm{br}^{2}/\mu\varepsilon E_c^{3}$.
These limits are mapped onto the low-noise amplifier, power amplifier,
and switch/phase-shifter functions of a transmit/receive front end and
quantified by a MATLAB device-physics model comparing GaN, GaAs, and Si
up to \SI{90}{GHz}. The purpose of this framework and its simulations
is to identify which material platform offers the best performance at
millimeter-wave frequency signals.
\end{abstract}

% ======================================================================
\section{Introduction}
% ======================================================================
Early microwave systems adopted the \emph{monolithic microwave
integrated circuit} (MMIC)---an \emph{integration} technology, not a
material. The active material underneath that technology moved from
silicon to GaAs (MESFETs, then heterostructure FETs with InGaAs
channels) and now to GaN \cite{gan_mmic_mmwave,pengelly2012}. It is
tempting to attribute this progression to lattice structure, but the
decisive variables are the energy bandgap, the interface electrostatics
of the nitrides, and the kinetics of hot carriers; lattice mismatch is
in fact an \emph{obstacle} that GaN growth had to overcome, not a
driver of its adoption. In this paper every performance claim is
derived from the underlying physics---symmetry, electrostatics, and
carrier--phonon energy balance---with the literature cited only for
measured material constants and fabricated-device results.
Section~\ref{sec:physics} builds the material case,
Sec.~\ref{sec:circuits} maps it onto circuit blocks, and
Sec.~\ref{sec:sim} quantifies the GaN--GaAs--Si comparison with a
simulation model.

% ======================================================================
\section{Physics of the GaN Advantage}
\label{sec:physics}
% ======================================================================

\subsection{Crystal symmetry and built-in polarization}
GaN is tetrahedrally bonded but stacks its Ga--N bilayers in the
hexagonal ABAB sequence (the wurtzite structure, space group $P6_3mc$),
which possesses \emph{no inversion centre}; its point group $6mm$
therefore admits a nonzero macroscopic polarization along the $c$ axis
\cite{morkoc}. Because the bond is strongly ionic (Pauling
electronegativities 1.81 for Ga, 3.04 for N), each bilayer carries a
dipole moment, and the coherent sum over the crystal is a
\emph{spontaneous} polarization present with zero strain,
$P_\mathrm{SP}(\mathrm{GaN})=-0.034\,\si{C/m^2}$, increasing in
magnitude with Al content \cite{ambacher2000}:
\begin{equation}
P_\mathrm{SP}(x)=-0.090\,x-0.034\,(1-x)+0.021\,x(1-x)\ \ [\si{C/m^2}].
\label{eq:psp}
\end{equation}
A thin $\mathrm{Al}_x\mathrm{Ga}_{1-x}\mathrm{N}$ film grown coherently
on relaxed GaN is stretched to the GaN in-plane lattice constant,
$\varepsilon_\parallel=(a_\mathrm{GaN}-a(x))/a(x)$. The free surface
carries no vertical stress, so linear elasticity
($\sigma_{zz}=2C_{13}\varepsilon_\parallel+C_{33}\varepsilon_{zz}=0$)
fixes the out-of-plane strain, and the piezoelectric tensor converts
the strain state into an additional polarization:
\begin{equation}
\varepsilon_{zz}=-\frac{2C_{13}}{C_{33}}\,\varepsilon_\parallel,
\qquad
P_\mathrm{PZ}=2\varepsilon_\parallel\!\left(e_{31}-e_{33}\frac{C_{13}}{C_{33}}\right).
\label{eq:ppz}
\end{equation}
For tensile AlGaN on GaN both terms carry the same sign, so
$P_\mathrm{SP}$ and $P_\mathrm{PZ}$ add. The elastic and piezoelectric
constants entering Eq.~\eqref{eq:ppz} are measured quantities
\cite{ambacher1999,neuberger_iiiv}.

\subsection{Interface electrostatics and the confined channel}
\begin{figure}[t]
  \centering
  \includegraphics[width=0.92\linewidth]{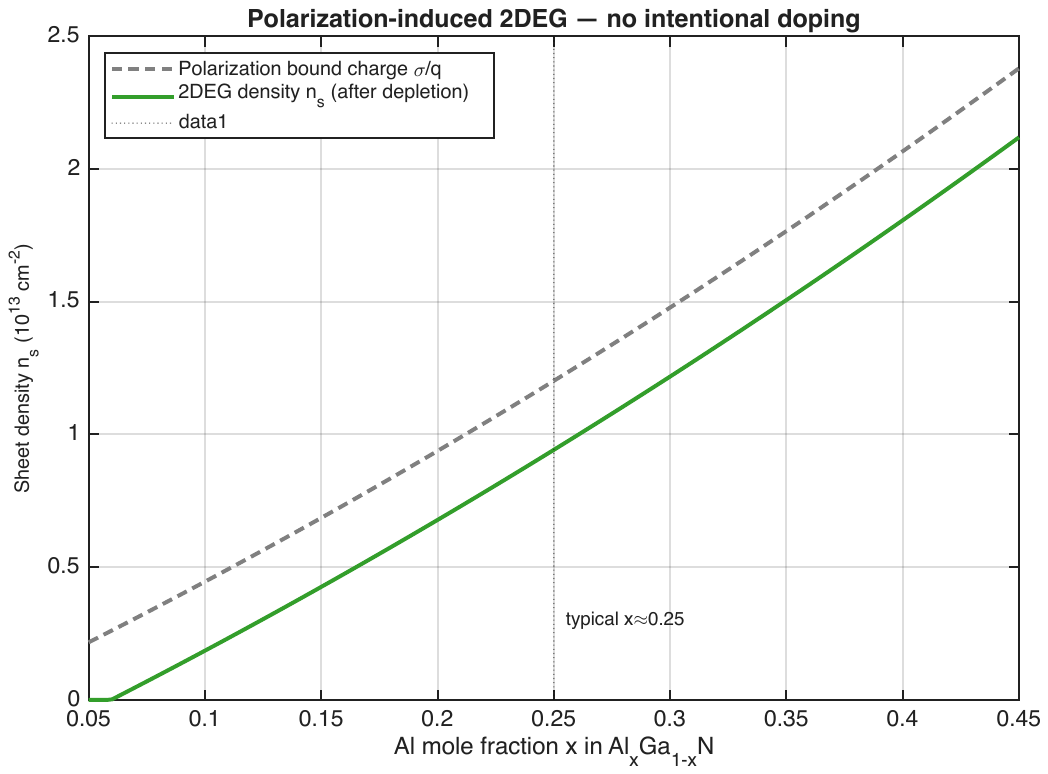}
  \caption{Simulated polarization bound charge and channel density
  versus Al mole fraction, computed from
  Eqs.~\eqref{eq:psp}--\eqref{eq:ns} with measured elastic and
  piezoelectric constants \cite{ambacher1999,ambacher2000}. Model
  result, not measured.}
  \label{fig:2deg}
\end{figure}

A spatially varying polarization is equivalent to a bound charge
$\rho_b=-\nabla\!\cdot\!\mathbf{P}$; at an abrupt AlGaN/GaN junction
this collapses to a fixed sheet
\begin{equation}
\sigma_\pi=\big|P_\mathrm{SP}^{\mathrm{AlGaN}}+P_\mathrm{PZ}
-P_\mathrm{SP}^{\mathrm{GaN}}\big| .
\label{eq:sigma}
\end{equation}
Integrating Poisson's equation across a barrier of thickness $d$ with a
Schottky boundary condition $q\phi_b$ at the surface gives the mobile
sheet density that accumulates against the interface,
\begin{equation}
n_s=\frac{\sigma_\pi}{q}
-\frac{\varepsilon_0\varepsilon_b}{d\,q^{2}}
\left[q\phi_b+E_F-\Delta E_C\right],
\label{eq:ns}
\end{equation}
with $\Delta E_C$ the conduction-band offset. For $x=0.25$ and
$d\approx\SI{20}{nm}$, Eqs.~\eqref{eq:psp}--\eqref{eq:ns} give
$n_s\approx1\times10^{13}\,\si{cm^{-2}}$ (Fig.~\ref{fig:2deg})---%
\emph{without any donor atoms}.

The electrons are held against the interface by their own image field,
$F_\mathrm{eff}\simeq qn_s/2\varepsilon\approx\SI{1}{MV/cm}$, an
approximately triangular potential well. Solving the Schr\"odinger
equation with a linear potential gives Airy-function bound states with
eigenvalues
\begin{equation}
E_n\simeq\left(\frac{\hbar^{2}}{2m^{*}}\right)^{\!1/3}
\left[\frac{3\pi}{2}\,qF_\mathrm{eff}\!\left(n+\frac{3}{4}\right)\right]^{2/3},
\label{eq:airy}
\end{equation}
with inter-subband spacings of order \SI{100}{meV}, so at room
temperature only the lowest subbands are occupied and the carrier
system is rigorously two-dimensional---a \emph{two-dimensional electron
gas} (2DEG) confined within $\sim$\SIrange{2}{3}{nm} of the interface.
In two dimensions the density of states is constant, $m^{*}/\pi\hbar^{2}$
per subband, so $E_F-E_0=\pi\hbar^{2}n_s/m^{*}\approx\SI{0.12}{eV}$ at
$n_s=10^{13}\,\si{cm^{-2}}$: the channel is a degenerate, metal-like
sheet ($E_F-E_0\gg k_BT=\SI{26}{meV}$). Two consequences matter for
circuits. First, the confining charge is the fixed lattice polarization
rather than ionised donors, so remote-impurity scattering is absent
\emph{by construction} and the sheet conductance $q\mu n_s$ is high
despite GaN's moderate bulk mobility. Second, $n_s$ is set by tensor
constants of the crystal, not by doping statistics, which favours
device-to-device uniformity and linearity.

\subsection{Pair-creation threshold and the breakdown field}
Avalanche breakdown begins when a carrier gains enough kinetic energy
to create an electron--hole pair. Momentum \emph{and} energy
conservation with parabolic, equal-mass bands set the threshold at
$E_\mathrm{th}\approx\tfrac{3}{2}E_g$: \SI{5.1}{eV} in GaN against
\SI{1.7}{eV} in Si. The probability that a carrier reaches
$E_\mathrm{th}$ ballistically over a mean free path $\lambda$ gives an
ionisation coefficient
$\alpha(E)\simeq\lambda^{-1}\exp[-E_\mathrm{th}/qE\lambda]$, and
breakdown occurs when the ionisation integral $\int\alpha\,dz$ reaches
unity. GaN suppresses $\alpha$ twice over: the threshold is
three times larger, \emph{and} $\lambda$ is shorter because the
polar optical phonon of GaN is unusually stiff
($\hbar\omega_\mathrm{LO}=\SI{92}{meV}$ versus \SI{36}{meV} in GaAs)
and couples strongly to hot electrons. The result is
$E_c\approx\SI{3.3}{MV/cm}$, an order of magnitude above Si and GaAs
(Table~\ref{tab:mat}). Because the threshold is set by the bandgap,
this ranking is fundamental---no process development can raise the
$E_c$ of Si or GaAs to the nitride value.

\subsection{Phonon-limited high-field transport}
At fields far beyond the ohmic regime, the power a carrier absorbs from
the field, $qEv$, is balanced by optical-phonon emission, and the drift
velocity saturates near
$v_\mathrm{sat}\sim\sqrt{\hbar\omega_\mathrm{LO}/m^{*}}$. For GaN
($m^{*}=0.2\,m_0$) this estimate gives $2.8\times10^{7}\,\si{cm/s}$,
in good agreement with the accepted $2.5\times10^{7}\,\si{cm/s}$
\cite{quay2008,palacios_mishra}. GaAs is \emph{not} limited by this
mechanism: its light $\Gamma$-valley electrons ($m^{*}=0.067\,m_0$)
transfer to heavy satellite valleys only \SI{0.29}{eV} up, so the
velocity peaks and then \emph{falls} with field. In GaN the satellite
valleys lie $\gtrsim\SI{1}{eV}$ above $\Gamma$, so no transfer occurs
below breakdown-level fields, and Si's multi-valley conduction band
saturates at barely $1\times10^{7}\,\si{cm/s}$. GaN is therefore the
only one of the three that sustains $\sim2.5\times10^{7}\,\si{cm/s}$
\emph{at} megavolt-per-centimetre fields---precisely the operating
point of a power device.

\subsection{Geometry-free device limits}
\begin{figure}[t]
  \centering
  \includegraphics[width=0.92\linewidth]{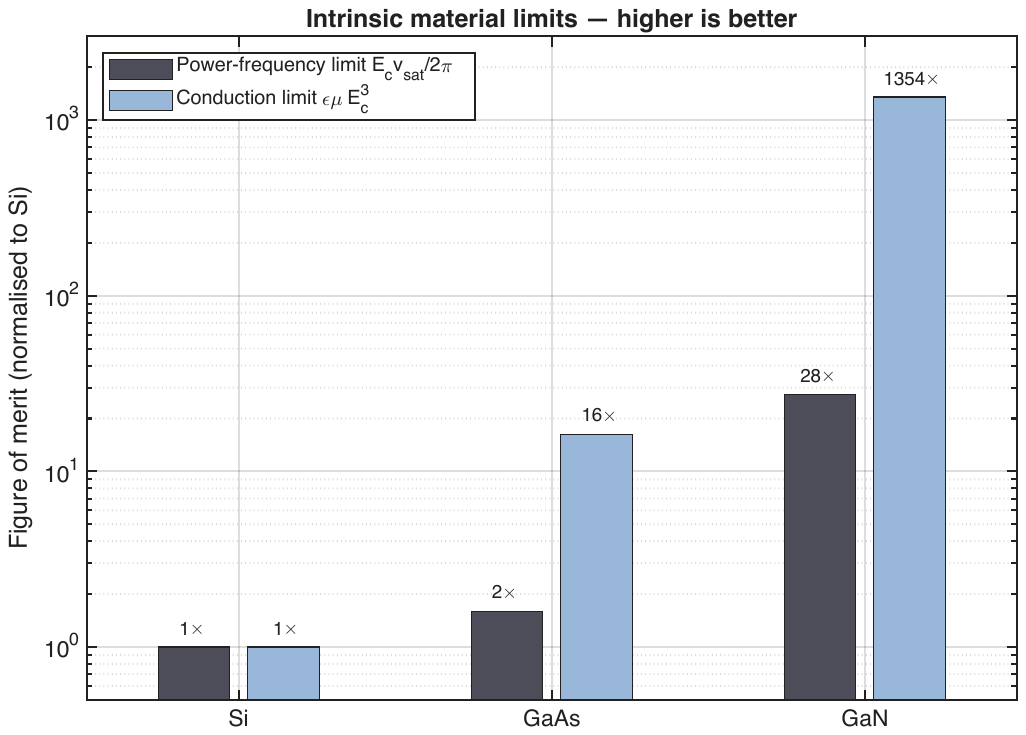}
  \caption{Simulated intrinsic material limits normalised to Si: the
  power--frequency product of Eq.~\eqref{eq:jfom} and the conduction
  limit $\mu\varepsilon E_c^{3}$ of Eq.~\eqref{eq:bfom}
  (parameters of Table~\ref{tab:mat}). Higher is better.}
  \label{fig:fom}
\end{figure}

The same physical length $L$ that a carrier must transit also has to
withstand the drain voltage. With transit-limited gain cut-off
$f_T=v_\mathrm{sat}/2\pi L$ and breakdown-limited swing
$V_\mathrm{br}\simeq E_cL$, the geometry cancels:
\begin{equation}
V_\mathrm{br}\,f_T=\frac{E_c\,v_\mathrm{sat}}{2\pi},
\label{eq:jfom}
\end{equation}
a pure material constant---the attainable power--frequency envelope. It
is $\sim$28$\times$ larger for GaN than Si (Fig.~\ref{fig:fom}).
Similarly, a unipolar drift region designed to hold $V_\mathrm{br}$
(triangular field profile, $L=2V_\mathrm{br}/E_c$, charge set by
Gauss's law) has a specific on-resistance
\begin{equation}
R_\mathrm{on}^\mathrm{sp}=\frac{L}{q\mu N_D}
=\frac{4V_\mathrm{br}^{2}}{\mu\,\varepsilon\,E_c^{3}},
\label{eq:bfom}
\end{equation}
so conduction loss at fixed blocking voltage improves as $E_c^{3}$:
three orders of magnitude in favour of GaN. Equations
\eqref{eq:jfom}--\eqref{eq:bfom} are the two limits plotted in
Fig.~\ref{fig:fom}; they were first tabulated across materials in
\cite{johnson1965,baliga1989}.

\begin{table}[t]
\centering
\caption{Representative room-temperature material parameters
(measured values \cite{neuberger_iiiv,quay2008}; GaN row is the
GaN-on-SiC platform).}
\label{tab:mat}
\footnotesize
\setlength{\tabcolsep}{4pt}
\begin{tabular}{lccc}
\toprule
Property & Si & GaAs & GaN \\
\midrule
$E_g$ (\si{eV})                          & 1.12 & 1.42 & 3.40 \\
$E_c$ (\si{MV/cm})                       & 0.30 & 0.40 & 3.30 \\
$v_\mathrm{sat}$ ($10^7$\,\si{cm/s})     & 1.0  & 1.2  & 2.5 \\
$\mu$ (\si{cm^2/V.s})                    & 1350 & 8500 & 1500--2000 \\
$\varepsilon_r$                          & 11.8 & 12.9 & 9.0 \\
$\hbar\omega_\mathrm{LO}$ (\si{meV})     & 63$^{\ddagger}$ & 36 & 92 \\
$\kappa$ (\si{W/cm.K})                   & 1.5  & 0.5  & 1.5$^{\dagger}$ \\
\bottomrule
\end{tabular}
\\[2pt]{\footnotesize $^{\dagger}$GaN layer; SiC substrate
\SIrange{3}{4}{W/cm.K}. $^{\ddagger}$Non-polar optical/intervalley.}
\end{table}

The one genuine weakness follows from the same physics: a large
$P\!\times\!f$ product concentrates dissipation in a nanometre-scale
channel. The dominant RF platform is therefore GaN-on-SiC, whose
substrate thermal conductivity (\SIrange{3}{4}{W/cm.K}) is roughly
eight times that of GaAs and preserves the material advantage under
continuous drive \cite{pengelly2012}.

% ======================================================================
\section{GaN in the RF Front-End}
\label{sec:circuits}
% ======================================================================
A transmit/receive (T/R) module requires three functions: low-noise
amplification (LNA), high-power amplification (HPA), and signal routing
and phase control (switches, phase shifters). We treat each with the
material limits of Sec.~\ref{sec:physics}.

\subsection{Receiver: noise from two-port first principles}
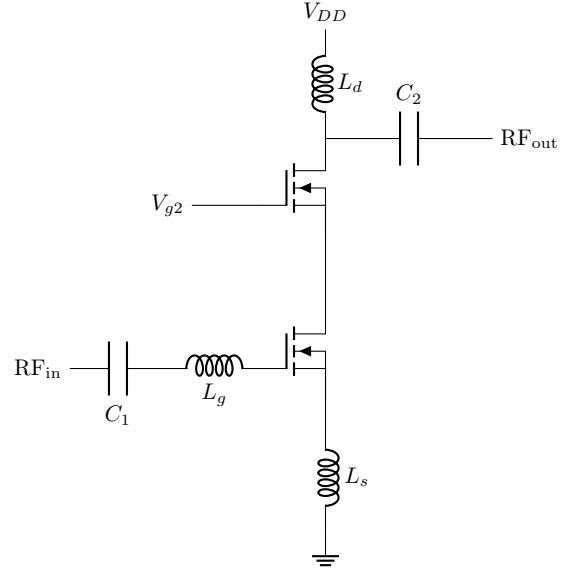
\begin{figure}[t]
\centering
\begin{circuitikz}[scale=0.85,transform shape]
  \draw (0,0) node[left]{RF$_\mathrm{in}$}
        to[C,l_=$C_1$] (1.5,0)
        to[L,l_=$L_g$] (3,0);
  % common-source HEMT (input device)
  \draw (3,0) node[nigfete,anchor=G](m1){};
  \draw (m1.S) -- ++(0,-0.4) to[L,l=$L_s$] ++(0,-1.6) node[ground]{};
  % cascode common-gate HEMT (stacked directly above m1)
  \draw (m1.D) -- ++(0,1.0) coordinate(cas);
  \draw (cas) node[nigfete,anchor=S](m2){};
  \draw (m2.G) -- ++(-1.1,0) node[left]{$V_{g2}$};
  % output bias tee: L_d choke to V_DD, C_2 DC block to RF_out
  \draw (m2.D) to[L,l_=$L_d$] ++(0,1.7) node[above]{$V_{DD}$};
  \draw (m2.D) to[C,l=$C_2$] ++(2.6,0) node[right]{RF$_\mathrm{out}$};
\end{circuitikz}
\caption{Cascode GaN HEMT LNA: inductive source degeneration ($L_s$)
sets a simultaneous noise/impedance match at $50\,\Omega$; the
common-gate device adds gain and reverse isolation
\cite{gan_reconfig_lna}.}
\label{fig:lna}
\end{figure}

All noise of a field-effect transistor can be referred to its input as
the thermal noise of the parasitic gate and source resistances,
$S_v=4k_BT(R_g+R_s)$, plus the channel current noise
$S_{i_d}=4k_BT\gamma g_m$, where $\gamma$ is the channel noise factor
of the hot-electron gas. Minimising the noise figure of this two-port
over the source impedance yields
\begin{equation}
F_\mathrm{min}\simeq1+2\,\frac{f}{f_T}\sqrt{\gamma\,g_m(R_g+R_s)},
\qquad f_T=\frac{g_m}{2\pi C_{gs}},
\label{eq:nfmin}
\end{equation}
i.e., noise grows linearly with $f/f_T$; the widely used one-constant
empirical form is recovered with $K=2\sqrt{\gamma}$ \cite{fukui1979}.
Equation~\eqref{eq:nfmin} says GaAs heterostructure FETs, with the
lightest electrons and hence the highest $f_T$ and smallest access
resistances, retain the lowest \emph{raw} $F_\mathrm{min}$; GaN sits
within about a decibel of it from X-band to W-band
\cite{gan_xband_lna_fab,gan_kuband_lna,gan_wband_lna}. GaN's receiver
advantage is instead set by \emph{survivability}. Any passive
protection loss $L$ placed ahead of the first transistor multiplies the
receiver noise factor, $F_\mathrm{rx}=L\,F_\mathrm{LNA}$: every decibel
of limiter loss is a decibel of noise figure. Because the GaN gate
withstands order-of-magnitude larger input power and recovers from
overload in nanoseconds \cite{gan_robust_lna,gan_reconfig_lna}, the
limiter can be removed---and with it its loss---so the \emph{system}
noise figure of a GaN receiver equals or beats the GaAs one in any
scenario with transmit leakage or jamming. Figure~\ref{fig:lna} shows
the cascode topology used in such designs: source inductance $L_s$
realises a $50\,\Omega$ noise/impedance match simultaneously, while the
common-gate device raises gain and reverse isolation
\cite{gan_reconfig_lna}.

\subsection{Transmitter: load-line power and waveform efficiency}
On an optimum resistive load line swung between the knee voltage
$V_\mathrm{knee}$ and $V_\mathrm{br}$, the fundamental output power is
\begin{equation}
P_\mathrm{out}=\tfrac{1}{8}\,I_\mathrm{max}\,
(V_\mathrm{br}-V_\mathrm{knee}),
\label{eq:pout}
\end{equation}
so breakdown voltage converts \emph{directly} into watts: this is
Eq.~\eqref{eq:jfom} at work, and it yields
\SIrange{5}{40}{W/mm} of gate periphery in GaN against
$\sim\SI{1}{W/mm}$ in GaAs \cite{pengelly2012,gan_100w_matched}. The
large voltage swing also raises the optimum load resistance
$R_\mathrm{opt}=(V_\mathrm{br}-V_\mathrm{knee})/I_\mathrm{max}$ towards
$50\,\Omega$, easing broadband matching. Efficiency follows from
waveform overlap: for half-sinusoidal current,
\begin{equation}
\eta_D=\frac{\pi}{4}\,
\frac{V_\mathrm{br}-V_\mathrm{knee}}{V_\mathrm{br}+V_\mathrm{knee}},
\label{eq:etab}
\end{equation}
and shaping the voltage towards a square wave with harmonic
terminations removes the residual overlap; the power-added efficiency
$\mathrm{PAE}=(P_\mathrm{out}-P_\mathrm{in})/P_\mathrm{DC}
=\eta_D(1-1/G)$ reaches 57\% at X-band with internal harmonic control
\cite{gan_harmonic_pa} and remains useful even at E-band
\cite{gan_eband_pa} and in pulsed L-band transmitters
\cite{gan_lband_tx,meneghesso2018}.

\subsection{Routing: switches, phase shifters, one-chip T/R}
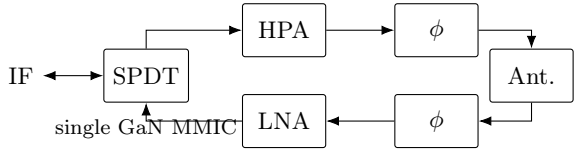
\begin{figure}[t]
\centering
\begin{tikzpicture}[
  node distance=6mm and 7mm,
  blk/.style={draw,minimum height=7mm,minimum width=11mm,font=\small,rounded corners=1pt},
  >=Latex]
  \node[blk] (sw) {SPDT};
  \node[blk,right=of sw,yshift=6mm] (hpa) {HPA};
  \node[blk,right=of sw,yshift=-6mm] (lna) {LNA};
  \node[blk,right=9mm of hpa] (ps1) {$\phi$};
  \node[blk,right=9mm of lna] (ps2) {$\phi$};
  \node[blk,right=7mm of $(ps1)!0.5!(ps2)$] (ant) {Ant.};
  \draw[<->] (sw.west) -- ++(-8mm,0) node[left,font=\small]{IF};
  \draw[->] (sw) |- (hpa);
  \draw[<-] (sw) |- (lna);
  \draw[->] (hpa) -- (ps1);
  \draw[<-] (lna) -- (ps2);
  \draw[->] (ps1) -| (ant);
  \draw[<-] (ps2) -| (ant);
  \node[font=\footnotesize,align=center,below=1mm of sw]{single GaN MMIC};
\end{tikzpicture}
\caption{One GaN process serves HPA, LNA, phase shifters ($\phi$), and
the T/R switch; the overload-tolerant receiver relaxes the need for a
circulator/limiter.}
\label{fig:tr}
\end{figure}

A series--shunt FET switch or switched phase-shifter bit is bounded by
the cut-off $f_\mathrm{co}=1/2\pi R_\mathrm{on}C_\mathrm{off}$. Here
the polarization physics returns: $R_\mathrm{on}\propto1/q\mu n_s$, and
the doping-free $10^{13}\,\si{cm^{-2}}$ channel drives it down, while
the modest permittivity of GaN keeps $C_\mathrm{off}$ small---high
$f_\mathrm{co}$ \emph{and} kilowatt-class power handling in the same
element \cite{gan_hemts_ch14}. The system consequence
(Fig.~\ref{fig:tr}) is that LNA, HPA, and switch share one process, so
the entire T/R function integrates on a single die, and the robust
receiver relaxes the circulator/limiter that GaAs front ends require
\cite{pengelly2012,tr_module_ref}.

% ---- T/R module block diagram (TikZ) ----

% ---- Cascode LNA schematic (circuitikz) ----

% ======================================================================
\section{Simulated Comparison: GaN vs GaAs vs Si}
\label{sec:sim}
% ======================================================================
\subsection{Methodology and disclaimer}
The figures in this section are produced by a compact device-physics
model (\texttt{figures/generate\_all\_figures.m}, MATLAB/Octave) with
the parameters of Table~\ref{tab:mat}. They are \emph{simulated
measurements}---model outputs cast in the units an instrument would
report---and are not laboratory data. Pulse edges are modelled as
first-order gain-bandwidth responses ($\tau\propto1/f_T$); pulse-top
droop uses a single-pole thermal response
$\Delta T(t)=P_\mathrm{diss}R_{th}(1-e^{-t/\tau_{th}})$; noise uses
Eq.~\eqref{eq:nfmin} with $\gamma=0.64$ (i.e., $K=1.6$).

\subsection{Pulsed-RF edge speed and thermal droop}
\begin{figure}[t]
  \centering
  \includegraphics[width=\linewidth]{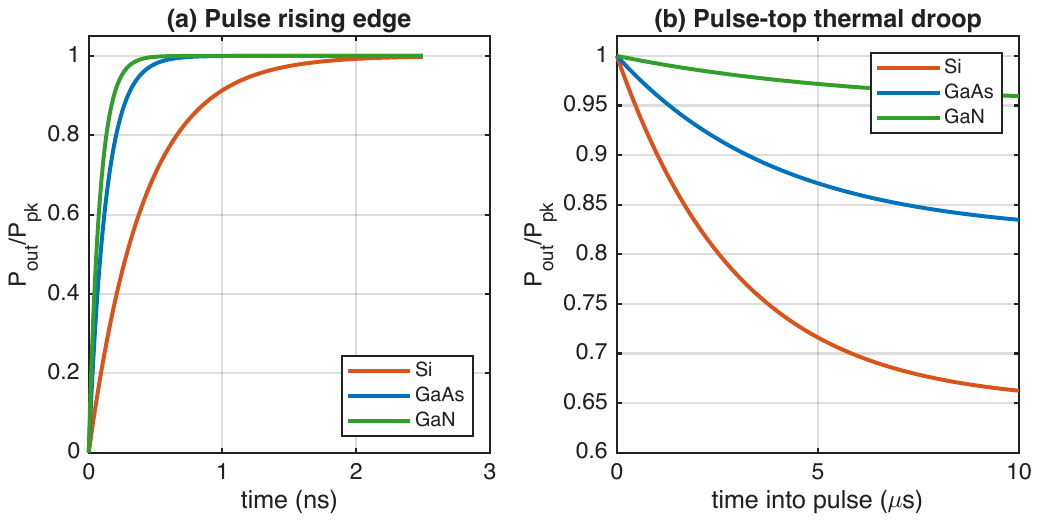}
  \caption{Simulated pulsed-RF envelope. (a) Rising edge; (b) pulse-top
  thermal droop over a \SI{10}{\micro s} pulse. Model outputs, not
  measured.}
  \label{fig:pulse}
\end{figure}

Figure~\ref{fig:pulse} compares (a) the rising edge and (b) the
pulse-top droop of a high-power RF pulse. GaN pairs the fastest edge
(largest usable gain-bandwidth at the required voltage swing) with the
flattest top, because the SiC heat path suppresses junction heating
over the pulse; Si is both slow and droops heavily, GaAs is
intermediate. For radar, a fast flat pulse translates directly into
range resolution and low spectral regrowth \cite{gan_lband_tx}.

\subsection{Minimum noise figure to 90\,GHz}
\begin{figure}[t]
  \centering
  \includegraphics[width=0.92\linewidth]{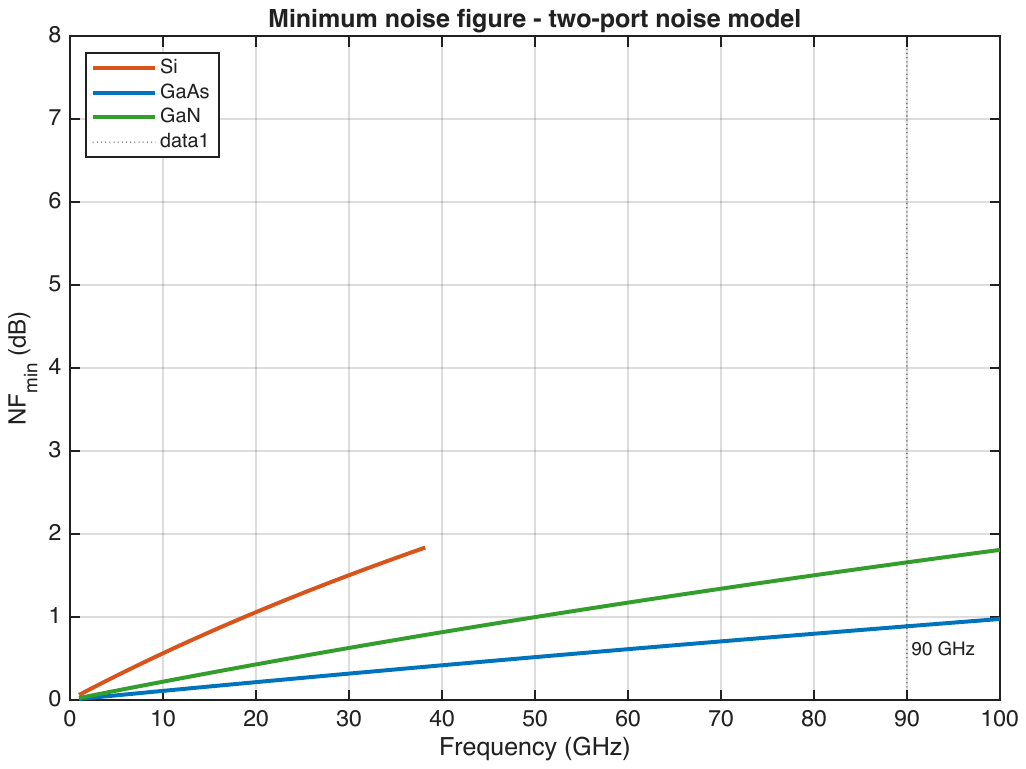}
  \caption{Simulated minimum noise figure versus frequency from the
  two-port model of Eq.~\eqref{eq:nfmin}. Si is truncated near its
  $f_T$, where the model---and the device---cease to be useful. Model
  outputs, not measured.}
  \label{fig:nf}
\end{figure}

Figure~\ref{fig:nf} evaluates Eq.~\eqref{eq:nfmin} to \SI{100}{GHz}.
As derived in Sec.~3.1, GaAs holds the lowest raw noise floor, GaN
tracks it within about \SI{1}{dB}, and the Si curve diverges as
$f\!\to\!f_T$: a physics-level statement that Si cannot serve as a
practical \SI{90}{GHz} LNA. The engineering trade is then explicit:
GaAs buys the last decibel of sensitivity; GaN buys survivability at
nearly the same sensitivity---and wins outright once limiter loss is
counted.

% ======================================================================
\section{Discussion}
% ======================================================================
Three system-level consequences follow from the device physics.
\emph{(i) Limiter elimination:} an overload-tolerant first stage
removes a loss element whose noise contribution is irreducible by any
later gain. \emph{(ii) Single-process integration:} one GaN MMIC
carries the full T/R function, shrinking phased-array element cost
\cite{pengelly2012,tr_module_ref}. \emph{(iii) The doping-free
channel:} because carriers are supplied by lattice polarization rather
than implanted or epitaxial donors, ion implantation in GaN is reserved
for isolation and selective-area contacts rather than channel
formation \cite{zolper1997,usman2009,pearton_process}, and the channel
is intrinsically uniform, linear, and radiation-hard. A natural
extension of this work is an electro-thermal co-simulation quantifying
how GaN-on-SiC versus GaN-on-Si substrates shift the droop curves of
Fig.~\ref{fig:pulse}.

% ======================================================================
\section{Conclusion}
% ======================================================================
The Si$\,\to\,$GaAs$\,\to\,$GaN progression in RF ICs follows from a
closed chain of physics. A non-centrosymmetric hexagonal lattice
admits built-in polarization; elasticity and the piezoelectric tensor
turn a heterointerface into a fixed charge sheet; electrostatics and
quantum confinement convert that sheet into a degenerate two-dimensional
channel without doping. The bandgap sets the pair-creation threshold
and hence a tenfold breakdown-field advantage, while stiff polar
phonons and remote satellite valleys let GaN alone sustain
$2.5\times10^{7}\,\si{cm/s}$ at megavolt-per-centimetre fields. These
combine into geometry-free limits [Eqs.~\eqref{eq:jfom},
\eqref{eq:bfom}] that map one-to-one onto amplifier power
[Eq.~\eqref{eq:pout}], efficiency [Eq.~\eqref{eq:etab}], receiver
robustness, and switch cut-off---and the simulated GaN--GaAs--Si
comparisons make the ranking, and its honest exceptions, quantitative.

% ======================================================================
\section*{Acknowledgment}
% ======================================================================
The authors gratefully acknowledge Hrant Vardanyan of Microsoft, Surface
Team, for his
guidance and mentorship throughout the course of this work. His insight
into the underlying theoretical framework, and his encouragement to
direct this study specifically toward gallium nitride device physics
and its associated calculations, were instrumental in shaping both the
direction and the analytical depth of this paper. The authors sincerely
thank him for the time, technical guidance, and support that made this
work possible. The authors also thank ChatGPT (OpenAI) for its
assistance with Matlab simulation scripting and, in particular, for its
substantial help with \LaTeX{} document formatting.

\balance
\bibliographystyle{ieeetr}
{\footnotesize
\bibliography{references}

@article{johnson1965,
  author  = {Johnson, E. O.},
  title   = {Physical Limitations on Frequency and Power Parameters of Transistors},
  journal = {RCA Review},
  volume  = {26}, pages = {163--177}, year = {1965},
  note    = {\url{https://ieeexplore.ieee.org/document/1147520}}
}

@article{baliga1989,
  author  = {Baliga, B. J.},
  title   = {Power Semiconductor Device Figure of Merit for High-Frequency Applications},
  journal = {IEEE Electron Device Letters},
  volume  = {10}, pages = {455--457}, year = {1989},
  doi     = {10.1109/55.43098},
  note    = {\url{https://ieeexplore.ieee.org/document/43098/}}
}

@article{fukui1979,
  author  = {Fukui, H.},
  title   = {Optimal Noise Figure of Microwave {GaAs} {MESFET}s},
  journal = {IEEE Transactions on Electron Devices},
  volume  = {26}, number = {7}, pages = {1032--1037}, year = {1979},
  doi     = {10.1109/T-ED.1979.19541},
  note    = {\url{https://ieeexplore.ieee.org/document/1480119/}}
}

@article{ambacher1999,
  author  = {Ambacher, O. and Smart, J. and Shealy, J. R. and Weimann, N. G.
             and Chu, K. and Murphy, M. and Schaff, W. J. and Eastman, L. F.
             and Dimitrov, R. and Wittmer, L. and Stutzmann, M. and
             Rieger, W. and Hilsenbeck, J.},
  title   = {Two-Dimensional Electron Gases Induced by Spontaneous and
             Piezoelectric Polarization Charges in {N-} and {Ga-}face
             {AlGaN/GaN} Heterostructures},
  journal = {Journal of Applied Physics},
  volume  = {85}, number = {6}, pages = {3222--3233}, year = {1999},
  doi     = {10.1063/1.369664},
  note    = {\url{https://scispace.com/pdf/two-dimensional-electron-gases-induced-by-spontaneous-and-11pht7aexp.pdf}}
}

@article{ambacher2000,
  author  = {Ambacher, O. and Foutz, B. and Smart, J. and Shealy, J. R.
             and Weimann, N. G. and Chu, K. and Murphy, M. and
             Sierakowski, A. J. and Schaff, W. J. and Eastman, L. F.
             and Dimitrov, R. and Mitchell, A. and Stutzmann, M.},
  title   = {Two Dimensional Electron Gases Induced by Spontaneous and
             Piezoelectric Polarization in Undoped and Doped
             {AlGaN/GaN} Heterostructures},
  journal = {Journal of Applied Physics},
  volume  = {87}, number = {1}, pages = {334--344}, year = {2000},
  doi     = {10.1063/1.371866},
  note    = {\url{https://ui.adsabs.harvard.edu/abs/2000JAP....87..334A/abstract}}
}

@incollection{palacios_mishra,
  author    = {Palacios, T. and Mishra, U. K.},
  title     = {{GaN}-Based Transistors for High-Frequency Applications},
  booktitle = {Comprehensive Semiconductor Science and Technology, Sec.~5.06},
  publisher = {Elsevier}, year = {2011},
  note      = {\url{https://www.researchgate.net/publication/288207184_GaN-Based_Transistors_for_High-Frequency_Applications}}
}

@book{quay2008,
  author    = {Quay, R.},
  title     = {Gallium Nitride Electronics},
  publisher = {Springer}, series = {Springer Series in Materials Science},
  volume    = {96}, year = {2008},
  doi       = {10.1007/978-3-540-71892-5},
  note      = {\url{https://doi.org/10.1007/978-3-540-71892-5}}
}

@book{morkoc,
  author    = {Morko\c{c}, H.},
  title     = {Nitride Semiconductors and Devices},
  publisher = {Springer}, series = {Springer Series in Materials Science},
  volume    = {32}, address = {Berlin}, year = {1999},
  doi       = {10.1007/978-3-642-58562-3},
  note      = {\url{https://doi.org/10.1007/978-3-642-58562-3}}
}

@book{meneghesso2018,
  editor    = {Meneghesso, G. and Meneghini, M. and Zanoni, E.},
  title     = {Gallium Nitride-enabled High Frequency and High Efficiency Power Conversion},
  publisher = {Springer}, year = {2018},
  doi       = {10.1007/978-3-319-77994-2},
  note      = {\url{https://doi.org/10.1007/978-3-319-77994-2}}
}

@book{pearton_process,
  author    = {Pearton, S. J. and Abernathy, C. R. and Ren, F.},
  title     = {Gallium Nitride Processing for Electronics, Sensors and Spintronics},
  publisher = {Springer}, year = {2006},
  doi       = {10.1007/1-84628-359-0},
  note      = {\url{https://doi.org/10.1007/1-84628-359-0}}
}

@book{neuberger_iiiv,
  author    = {Neuberger, M.},
  title     = {{III--V} Semiconducting Compounds},
  publisher = {Electronic Properties Information Center, Hughes Aircraft Co.},
  address   = {Culver City, CA}, year = {1971}
}

@article{usman2009,
  author  = {Usman, M. and Nazir, A. and Aggerstam, T. and Linnarsson, M. K.
             and Hall\'{e}n, A.},
  title   = {Electrical and Structural Characterization of Ion Implanted {GaN}},
  journal = {Nuclear Instruments and Methods in Physics Research B},
  volume  = {267}, pages = {1561--1563}, year = {2009},
  doi     = {10.1016/j.nimb.2009.01.091},
  note    = {\url{https://doi.org/10.1016/j.nimb.2009.01.091}}
}

@article{zolper1997,
  author  = {Zolper, J. C.},
  title   = {Ion Implantation in Group {III}-Nitride Semiconductors:
             A Tool for Doping and Defect Studies},
  journal = {Journal of Crystal Growth},
  volume  = {178}, pages = {157--167}, year = {1997},
  doi     = {10.1016/S0022-0248(97)00076-6},
  note    = {\url{https://doi.org/10.1016/S0022-0248(97)00076-6}}
}

@article{pengelly2012,
  author  = {Pengelly, R. S. and Wood, S. M. and Milligan, J. W. and
             Sheppard, S. T. and Pribble, W. L.},
  title   = {A Review of {GaN} on {SiC} High Electron-Mobility Power
             Transistors and {MMIC}s},
  journal = {IEEE Transactions on Microwave Theory and Techniques},
  volume  = {60}, number = {6}, pages = {1764--1783}, year = {2012},
  doi     = {10.1109/TMTT.2012.2187535},
  note    = {\url{https://scispace.com/pdf/a-review-of-gan-on-sic-high-electron-mobility-power-4mjzeqemry.pdf}}
}

@inproceedings{gan_xband_lna_fab,
  author    = {Lei, Pang and Xiaojuan, Chen and Xinyu, Liu},
  title     = {An {AlGaN/GaN} {HEMT}-Based Monolithic Integrated X-band
               Low Noise Amplifier},
  booktitle = {Institute of Microelectronics, Chinese Academy of Sciences,
               Beijing (IEEE; exact conference name unconfirmed --
               see IEEE Xplore Document 6238209)},
  year      = {},
  note      = {\url{https://ieeexplore.ieee.org/document/6238209/}}
}

@inproceedings{gan_harmonic_pa,
  author    = {Yamanaka, K. and Morimoto, T. and Chaki, S. and
               Nakayama, M. and Hirano, Y.},
  title     = {X-band Internally Harmonic Controlled {GaN} {HEMT}
               Amplifier with 57\% Power Added Efficiency},
  booktitle = {Proceedings of the 6th European Microwave Integrated
               Circuits Conference (EuMIC)},
  address   = {Manchester, UK}, pages = {61--64}, year = {2011},
  note      = {\url{https://ieeexplore.ieee.org/document/6102788}}
}

@inproceedings{gan_wband_lna,
  author    = {Lardizabal, S. and Hwang, K. C. and Kotce, J. and
               Brown, A. and Fung, A.},
  title     = {Wideband {W}-Band {GaN} {LNA} {MMIC} with State-of-the-Art
               Noise Figure},
  booktitle = {IEEE Compound Semiconductor Integrated Circuit Symposium (CSICS)},
  year      = {2016}, doi = {10.1109/CSICS.2016.7751079},
  note      = {Invited; \url{https://doi.org/10.1109/CSICS.2016.7751079}}
}

@inproceedings{gan_100w_matched,
  author    = {Noto, H. and Maehara, H. and Uchida, H. and Koyanagi, M.
               and Utsumi, H. and Nishihara, J. and Otsuka, H. and
               Yamanaka, K. and Nakayama, M. and Hirano, Y.},
  title     = {X- and {Ku}-band Internally Matched {GaN} Amplifiers with
               More Than 100~W Output Power},
  booktitle = {Proceedings of the 7th European Microwave Integrated
               Circuits Conference (EuMIC)},
  address   = {Amsterdam, The Netherlands}, pages = {695--698}, year = {2012},
  note      = {\url{https://ieeexplore.ieee.org/document/6459316}}
}

@inproceedings{gan_robust_lna,
  author    = {Rudolph, M. and Dewitz, M. and Liero, A. and Khalil, I.
               and Chaturvedi, N. and Wipf, C. and Bertenburg, R. M.
               and Miller, J. and W\"urfl, J. and Heinrich, W. and
               Tr\"ankle, G.},
  title     = {Highly Robust X-Band {LNA} with Extremely Short Recovery Time},
  booktitle = {IEEE {MTT-S} International Microwave Symposium (IMS) Digest},
  pages     = {781--784}, year = {2009},
  note      = {\url{https://ieeexplore.ieee.org/document/5165813/}}
}

@inproceedings{gan_reconfig_lna,
  author    = {Kobayashi, K. W. and Campbell, C. and Lee, C. and
               Gallagher, J. and Shust, J. and Botelho, A.},
  title     = {A Reconfigurable {S}-/{X}-band {GaN} Cascode {LNA} {MMIC}},
  booktitle = {IEEE Compound Semiconductor Integrated Circuit Symposium (CSICS)},
  address   = {Miami, FL, USA}, year = {2017},
  note      = {\url{https://ieeexplore.ieee.org/document/8240424/}}
}

@inproceedings{gan_kuband_lna,
  author    = {Guo, Fengqiang and Yao, Zhihong},
  title     = {Design of A {Ku}-Band {AlGaN/GaN} Low Noise Amplifier},
  booktitle = {2014 3rd Asia-Pacific Conference on Antennas and
               Propagation (APCAP)},
  address   = {Harbin, China}, pages = {1406--1408}, year = {2014},
  doi       = {10.1109/APCAP.2014.6992789},
  note      = {\url{https://ieeexplore.ieee.org/abstract/document/6992789/}}
}

@inproceedings{gan_lband_tx,
  author    = {Garg, Samriti Kumar and Aich, Suman and Dhar, Jolly},
  title     = {{GaN} Based L-Band High Power and High Efficiency Pulsed
               Transmitter},
  booktitle = {2015 IEEE International Microwave and {RF} Conference (IMaRC)},
  address   = {Ahmedabad, India}, pages = {155--158}, year = {2015},
  note      = {\url{https://ieeexplore.ieee.org/document/7411438}}
}

@inproceedings{gan_eband_pa,
  author    = {Micovic, M. and Kurdoghlian, A. and Moyer, H. P. and
               Hashimoto, P. and Hu, M. and Antcliffe, M. and
               Willadsen, P. J. and Wong, W. S. and Bowen, R. and
               Milosavljevic, I. and Yoon, Y. and Schmitz, A. and
               Wetzel, M. and McGuire, C. and Hughes, B. and Chow, D. H.},
  title     = {{GaN} {MMIC} {PA}s for E-Band (71~{GHz}--95~{GHz}) Radio},
  booktitle = {IEEE Compound Semiconductor Integrated Circuit Symposium (CSIC)},
  address   = {Monterey, CA, USA}, year = {2008},
  note      = {\url{https://ieeexplore.ieee.org/document/4674462/}}
}

@inproceedings{gan_mmic_mmwave,
  author    = {Micovic, M. and Kurdoghlian, A. and Moyer, H. P. and
               Hashimoto, P. and Schmitz, A. and Milosavljevic, I. and
               Willadsen, P. J. and Wong, W.-S. and Duvall, J. and Hu, M.
               and Wetzel, M. and Chow, D. H.},
  title     = {{GaN} {MMIC} Technology for Microwave and Millimeter-wave
               Applications},
  booktitle = {IEEE Compound Semiconductor Integrated Circuit Symposium
               (CSIC) Digest},
  pages     = {173--176}, year = {2005},
  note      = {\url{https://ieeexplore.ieee.org/document/1531801}}
}

@incollection{gan_hemts_ch14,
  author    = {Ng, Geok Ing and Arulkumaran, Subramaniam},
  title     = {{GaN} {HEMT}s Technology and Applications (Chapter 14)},
  booktitle = {Nano-Semiconductors: Devices and Technology},
  editor    = {Iniewski, Krzysztof},
  publisher = {CRC Press / Routledge}, pages = {377--414}, year = {2011},
  doi       = {10.1201/b11388-18},
  note      = {\url{https://www.routledgehandbooks.com/doi/10.1201/b11388-18}}
}

@inproceedings{tr_module_ref,
  author    = {Whelan, Colin S. and Kolias, Nicholas J. and Brierley,
               Steven and MacDonald, Chris and Bernstein, Steven},
  title     = {{GaN} Technology for Radars},
  booktitle = {{CS} MANTECH Conference},
  address   = {Boston, MA, USA}, year = {2012},
  note      = {\url{https://csmantech.org/paper/gan-technology-for-radars/};
               PDF: \url{https://myreader.toile-libre.org/uploads/My_62d4ce79be28c.pdf}}
}
}

\end{document}